\documentstyle[11pt,aaspp4]{article}

\begin{document}

\title{The Dusty Starburst Nucleus of M33\footnotemark[1]}

\footnotetext[1]{Based on observations made with the NASA/ESA Hubble
Space Telescope, obtained from the data Archive at the Space Telescope
Science Institute, which is operated by the Association of
Universities for Research in Astronomy, Inc., under NASA contract
NAS5-26555.}

\author{Karl D.\ Gordon\altaffilmark{2,3}, 
   M.\ M.\ Hanson\altaffilmark{4,5,6}, 
   Geoffrey C.\ Clayton\altaffilmark{2,3,7}, 
   G.\ H.\ Rieke\altaffilmark{4}, and
   K.\ A.\ Misselt\altaffilmark{2,7}}
\altaffiltext{2}{Department of Physics \& Astronomy, Louisiana State
   University, Baton Rouge, LA 70803}
\altaffiltext{3}{Visiting Astronomer at the Infrared Telescope
Facility, which is operated by the University of Hawaii under contract
from the National Aeronautics and Space Administration}
\altaffiltext{4}{Steward Observatory, University of Arizona, Tucson,
   AZ 85721}
\altaffiltext{5}{present address: Department of Physics, University of
   Cincinnati, Cincinnati, OH 45221}
\altaffiltext{6}{Hubble Fellow}
\altaffiltext{7}{Visiting Astronomer, Kitt Peak National Observatory,
National Optical Astronomy Observatories, which is operated by the
Association of Universities for Research in Astronomy, Inc.\ (AURA)
under cooperative agreement with the National Science Foundation. }

\lefthead{Gordon et al.}
\righthead{Dusty M33 Nucleus}

\begin{abstract}
   We have thoroughly characterized the ultraviolet to near-infrared
(0.15 - 2.2 $\micron$) spectral energy distribution (SED) of the
central parsec of the M33 nucleus through new infrared photometry and
optical/near-infrared spectroscopy, combined with ultraviolet/optical
observations from the literature and the HST archive. The SED shows
evidence for a significant level of attenuation, which we model through
a Monte Carlo radiative transfer code as a shell of clumpy Milky Way-type
dust ($\tau_V \sim 2 \pm 1$). The discovery of Milky Way-type dust (with
a strong 2175~\AA\ bump) internal to the M33 nucleus is different from
previous work which has found SMC-like dust (no bump) near starburst
regions.  The amount by which dust can be processed may be related to
the mass and age of the starburst as well as the extent to which the
dust can shield itself. Our starburst models include the effects of
this dust and can fit the SED if the nucleus was the site of a
moderate ($\sim$$10^8~L_{\sun}$ at 10 Myrs) episode of coeval star
formation about 70 Myrs ago. This result is quite different from
previous studies which resorted to multiple stellar populations
(between 2 and 7) attenuated by either no or very little internal
dust. The M33 nuclear starburst is remarkably similar to an older
version (70 Myr versus 10 Myr) of the ultra-compact starburst in the
center of the Milky Way.
 
\end{abstract}


\section{Introduction}

   The nucleus of M33 (NGC 598, Triangulum galaxy) has been studied
extensively, but its true nature has remained elusive.  It is bright
and compact, has a FWHM of $0\farcs 16$ ($= 0.6$ pc) in HST F555W
observations, and lacks a substantial black hole (\cite{kor93};
\cite{mas96}; \cite{lau98}).  Van den Bergh (1991) reviews studies on the
stellar populations in M33, including the nucleus.  Previous work has
found that the nucleus is composed of roughly two stellar populations,
one with an age of $10^7 - 10^8$ years and one with an age of $10^9 -
10^{10}$ years (\cite{oco83}; \cite{cia84}; \cite{sch90}).

  Previous attempts at modeling the stellar population of the M33
nucleus have either ignored the effects of reddening internal to M33
or modeled it assuming a simple screen of dust. The presence of
ultraviolet (UV) (especially far-UV [\cite{mas96}]) flux from the M33
nucleus, and its small size, imply a coeval, young stellar
population. The flat UV/optical spectrum of the M33 nucleus [Figure~1c
of McQuade, Calzetti, \& Kinney (1995)], when coupled with the
assumption of a single young stellar population, can only be explained
by significant reddening by dust internal to M33 (\cite{cal94}).  The
work of Witt, Thronson, \& Capuano (1992) demonstrates the importance
of accurately treating the radiative transfer through the dust in
galaxies.

  The M33 nucleus is ideal for investigating the role of dust in
starburst galaxies.  Previous work on starburst galaxies found that
the extinction curve lacks a 2175~\AA\ bump (\cite{cal94};
\cite{gor97}).  The only other dust which does not have a 2175~\AA\
bump is found in the Small Magellanic Cloud (SMC).  Gordon \& Clayton
(1998) reinvestigated the UV extinction in the SMC and found evidence
to support the processing of dust near sites of active star formation.
Dust with weak 2175~\AA\ bumps is seen along some Galactic and Large
Magellanic Cloud (LMC) sightlines (e.g.\ \cite{car88}; \cite{mis98}).
It is possible that the lack of a 2175~\AA\ bump is the result of
processing the dust near sites of active star formation. The work of
Gordon et al.\ (1997) was based on an IUE selected sample of
starburst galaxies and due to the large aperture of IUE, the sample
was observationally biased toward intrinsically bright starbursts with
small amounts of dust.  We probe the M33 nucleus with a model including
stars, gas, and dust.  We find that its dust has a strong 2175~\AA\
bump, in contrast to the behavior of the starburst dust studied
previously (\cite{gor97}).  This paper represents the beginning of a
series of papers devoted to understanding the role of dust on the
interpretation of stellar populations in starburst galaxies.

  In this paper, we find that the M33 SED can be modeled as a modest
($\sim$$7 \times 10^5~M_{\sun}$) $\sim$70 Myr old starburst enshrouded
by a shell of MW-like dust ($\tau_V \sim 2$) by using a starburst
model which includes stars, gas, and dust.  The 
indication of a young, very compact, and moderate luminosity starburst
in the central pc of M33 is interesting because of the possible
similarity to the very young and massive stars within the central 0.5
pc of the Milky Way. Because of their modest outputs and small sizes,
such systems would not stand out clearly in the nuclei of galaxies
outside the Local Group. The presence of two similar events in Local
Group galaxies suggests that such starbursts may be relatively common
and hence may play an important role in the evolution of the central
parsec-sized regions of galaxies.

\section{Data}

  The observations of the M33 nucleus can be divided into two groups;
photometry and spectroscopy.  For photometric measurements, the
background level is determined using an annulus surrounding the
nucleus.  As a result, the photometry of the M33 nucleus does not
include a significant contribution from the disk or bulge of M33.  The
background levels for spectroscopic observations are usually measured
a significant distance from the nucleus.  This means they include
contributions from regions surrounding the nucleus, which
makes studying just the nucleus difficult. Thus, to isolate the
spectral behavior of the nucleus requires a judicious combination of
both photometry and spectroscopy.

\subsection{Photometry \label{sec_phot_obs}}

  The Ultraviolet Imaging Telescope (UIT, \cite{ste97}) observed M33
during the Astro-1 mission in the far-UV (B1 filter, $\lambda_c
\approx 1500$~\AA) and the near-UV (A1 filter, $\lambda_c \approx
2400$~\AA).  The UIT images, along with ground-based UBV images, were
analyzed by Massey et al.\ (1996).  They present UIT-B1, UIT-A1, U, B,
and V measurements of the nucleus of M33 (see Table~\ref{tab_obs_sed})
and note the nucleus is very red compared to the other sources in
their images.  Additional ground-based measurements at B and R are
given by Kormendy \& McClure (1993).  The zero points for the U, B, V,
and R magnitudes were taken from Bessell, Castelli, \& Plez (1998).

\begin{deluxetable}{lcccc}
\tablewidth{0pt}
\tablecaption{Photometric Observations \label{tab_obs_sed}}
\tablehead{\colhead{Filter} & 
           \colhead{$\lambda_{\rm eff}$\tablenotemark{a}} &
           \colhead{$\Delta\lambda_{\rm eq}$\tablenotemark{b}} &
           \colhead{$F_{\lambda}$} & \colhead{References} \\ 
           & \colhead{[\AA]} & \colhead{[\AA]} & 
           \colhead{[$10^{-15}$ ergs cm$^{-2}$ s$^{-1}$ \AA$^{-1}$]} & }
\startdata
UIT-B1 & 1534  &  355 & $4.88 \pm 0.73$ & 1 \nl
F160BW & 1554  &  575 & $6.33 \pm 0.95$ & 2 \nl
F170W  & 1695  &  520 & $6.03 \pm 0.90$ & 2 \nl
UIT-A1 & 2607  & 1147 & $2.34 \pm 0.35$ & 1 \nl
F300W  & 3045  &  879 & $5.12 \pm 0.51$ & 2 \nl
F336W  & 3349  &  495 & $5.36 \pm 0.54$ & 2 \nl
U      & 3641  &  641 & $4.41 \pm 0.44$ & 1 \nl
B      & 4421  &  960 & $9.31 \pm 0.93$ & 1 \nl
B      & 4421  &  960 & $9.39 \pm 0.66$ & 3 \nl
F555W  & 5314  & 1578 & $10.0 \pm 1.00$ & 2 \nl
V      & 5507  &  984 & $8.87 \pm 0.89$ & 1 \nl
R      & 6563  & 1591 & $10.3 \pm 1.03$ & 3 \nl
F814W  & 8365  & 2504 & $7.86 \pm 0.79$ & 2 \nl
F1042M & 10444 &  854 & $7.01 \pm 1.05$ & 2 \nl
J      & 12302 & 2036 & $5.89 \pm 0.59$ & 2 \nl
H      & 16379 & 2863 & $3.80 \pm 0.38$ & 2 \nl
K'     & 21201 & 2882 & $1.86 \pm 0.19$ & 2 \nl
\enddata
\tablenotetext{a}{$\lambda_{\rm eff}$ was calculated by using the
Set A PEGASE best fit model SED (see \S~\ref{sec_fits})}
\tablenotetext{b}{$\Delta\lambda_{\rm eq}$ was calculated by using
eq. 5 of \cite{gor98b}}
\tablerefs{(1) \cite{mas96}; (2) this work; (3) \cite{kor93}}
\end{deluxetable}

  Numerous WFPC2 images of the central region of M33 exist in the HST
archive.  We chose post-COSTAR images taken with at least two images
in each filter to facilitate the removal of cosmic rays.  The names,
filters, and exposure times of the images used are
listed in Table~\ref{tab_wfpc2_obs}.  All but the F170W and F336W
images were taken with the nucleus centered on the PC chip.  For each
filter, we combined the available images to create a cosmic ray free image
without saturated pixels.  The total counts from the nucleus in each
filter were calculated by summing the counts out to a radius $4\farcs
55$ after subtracting the background counts, which were determined
using an annulus with radii between $4\farcs 55$ and $6\farcs 83$.
It was necessary to use an aperture at least $4\arcsec$ in radius due
to the broad wings on the HST point spread function (\cite{hol95}).
These counts were converted to the fluxes presented 
in Table~\ref{tab_obs_sed} using the algorithms presented by Voit et
al.\ (1997) and Whitmore, Heyer, \& Baggett (1996).  The uncertainties
were assumed to be 15\% for F160BW, F170W, and F1042M images.  All
other images were assumed to have an uncertainty of 10\%.  These
values reflect the uncertainties in both the absolute calibration
and determination of the background level.

\begin{deluxetable}{ccccc}
\tablewidth{0pt}
\tablecaption{WFPC2 Images \label{tab_wfpc2_obs}}
\tablehead{\colhead{Image} & \colhead{Filter} &
             \colhead{Exp.\ Time} \\ 
           \colhead{Name} & & \colhead{[sec]} }
\startdata
U3MR0101M & F160BW & 1200 \nl
U3MR0102M & F160BW & 1200 \nl
U2YE0207T & F170W & 900 \nl
U2YE0209T & F170W & 900 \nl
U3MR0103M & F300W & 800 \nl
U3MR0104M & F300W & 800 \nl
U2YE020BT & F336W & 900 \nl
U2YE020DT & F336W & 900 \nl
U2E20507T & F555W & 400 \nl
U2E20508T & F555W & 400 \nl
U2E20509T & F555W & 400 \nl
U2E2050AT & F555W & 400 \nl
U2E2050BT & F555W & 40 \nl
U2E2050CT & F555W & 40 \nl
U2E2050DT & F814W & 400 \nl
U2E2050ET & F814W & 400 \nl
U2E2050FT & F814W & 400 \nl
U2E2050GT & F814W & 40 \nl
U2E2050HT & F814W & 40 \nl
U2E20501T & F1042M & 500 \nl
U2E20502T & F1042M & 500 \nl
U2E20503T & F1042M & 700 \nl
U2E20504T & F1042M & 300 \nl
U2E20505T & F1042M & 500 \nl
U2E20506T & F1042M & 500 \nl
\enddata
\end{deluxetable}

  Near infrared images at $J$, $H$, and $K'$ of the central region of
M33 were taken on 1998 March 2 (UT) using the NSFCAM instrument on
NASA's Infrared Telescope Facility (IRTF).  The detector was a
$256 \times 256$ InSb array with an image scale of $0\farcs 3$/pixel
giving a field-of-view of $76\farcs 3$.  For each filter, four images
were taken alternating with four sky exposures.  The sky images were
combined using a sigma clipping algorithm giving a final sky image
without stars.  The flatfield image was calculated by subtracting a
dark image from the sky image and normalizing the result.  Each of the
images of M33 was sky subtracted and flat fielded before being
combined using the sigma clipping algorithm to remove cosmic rays and
bad pixels.  The calibration of the images was done using standard
stars from Hunt et al.\ (1998) chosen to bracket $(J-K)$ colors
between -0.15 and 0.63.  The standard stars were observed throughout
the night bracketing airmasses between 1.0 and 2.2.  The $K$
magnitudes by Hunt et al.\ (1998) were converted to $K'$ magnitudes
using $K' = K + 0.2(H-K)$ (\cite{wai92}).  Total exposure times were
80 seconds for each of the three filters.  The $J$, $H$, and $K'$
magnitudes of the nucleus were extracted using a sky annulus with
radii between 40 and 60 pixels.  The magnitudes were transformed from
the CIT system to the homogeneous system defined by Bessell \& Brett
(1988).  The zero points for the $J$, $H$, and $K'$ magnitudes
were taken from Bessell et al.\ (1998) and the resulting fluxes are
given in Table~\ref{tab_obs_sed}.  These fluxes are consistent with
those given by Gallagher et al.\ (1982) within the uncertainties.

\subsection{Spectroscopy}

  International Ultraviolet Explorer (IUE) observations provide the
only ultraviolet spectra of the M33 nucleus.  The observations used
the large aperture ($10\arcsec \times 20\arcsec$).  Examining the
line-by-line (SILO) spectra of the two IUE observations with similar
position angles (SWP 52142 and LWP 1584), we found two sources were
present.  Using the MGEX and NEWCALIB routines provided in the
IUEIDL package (\cite{nic94}), we extracted and calibrated the two
spectra.  We imposed the IUE aperture on the UIT images of M33 and
confirmed that one of the spectra was from the nucleus and one was from a
nearby star (UIT 219, $d = 10\arcsec$, \cite{mas96}).  The integrated
flux from the UIT images (FUV0496 and NUV0402) in the IUE aperture
matched that of the sum of the two spectra within the uncertainties.

  Optical spectroscopy of the M33 nucleus was obtained using the
GoldCam CCD spectrometer on the 2.1--meter telescope at Kitt Peak
National Observatory on 1998 June 29 (UT).  Grating \#201 was used with a
slit width of 240~$\mu$m (3$\arcsec$) resulting in a resolution of
19~\AA\ FWHM over a spectral range from 3300--9100~\AA.  In the
spatial direction, the slit extended $5\farcm 2$ with a scale of
0.78$\arcsec$~pix$^{-1}$.  The M33 spectrum was obtained in a single
900~sec exposure with a position angle of 100$\arcdeg$ and at an
airmass of 1.39.  The flat field used in reducing this spectrum was
constructed from exposures of a quartz lamp corrected for the slit
illumination function using twilight sky exposures.  The spectrum was
wavelength calibrated using HeNeAr lamp exposures taken throughout
the night.  The rms deviation of the adopted solution about the
comparison line wavelengths was $\le 1.0$~\AA.  Atmospheric H$_2$O and
O$_2$ bands were removed using observations of hot stars and the method
described by Bica \& Alloin (1987).  A mean sky frame was constructed
from several exposures of galaxies that did not fill the spatial
direction on the chip.  This frame was subsequently scaled to match
the night sky lines in the M33 spectrum and subtracted from the M33
image.  A 1--dimensional spectrum 
was extracted using a 10$\arcsec$ aperture centered on the M33 nuclear
spectrum.  The non-nuclear contribution to the spectrum was subtracted
by fitting a 5th degree polynomial to the galactic background along the
spatial direction.  The resulting spectrum was flux calibrated using
several observations of spectrophotometric standards and the mean
extinction coefficients appropriate for KPNO.

\begin{figure}[tbp]
\begin{center}
\plotone{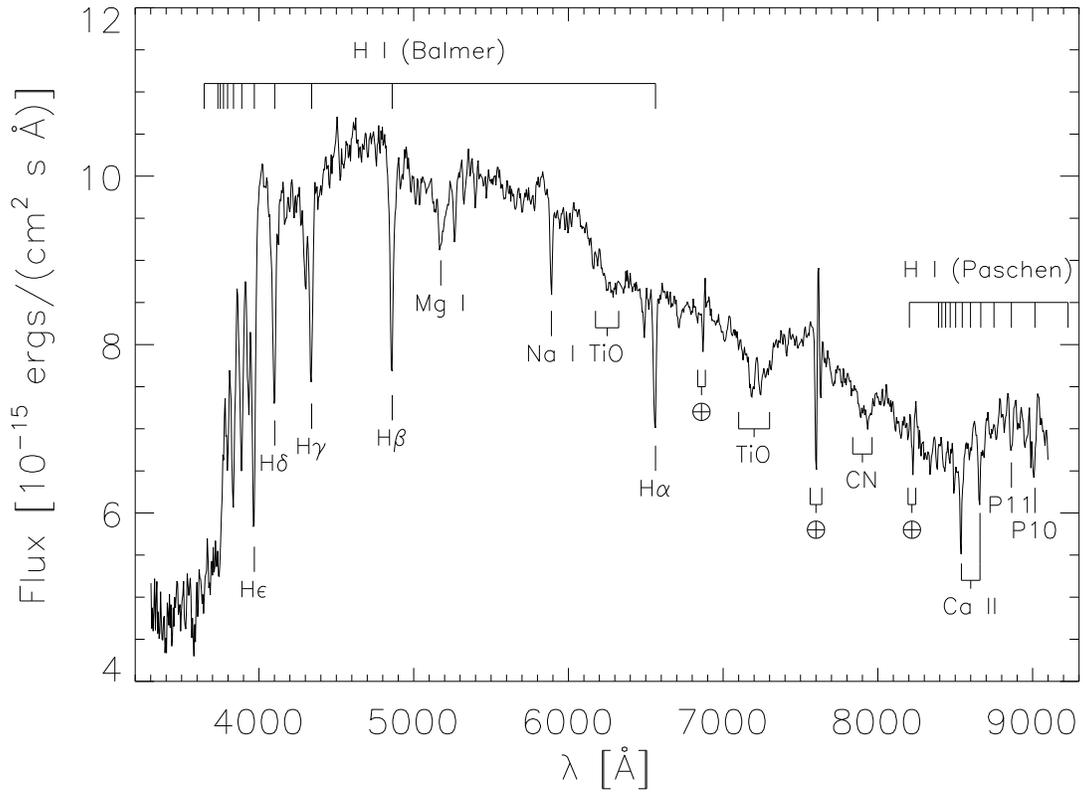}
\caption{The optical spectrum of the M33 nucleus is plotted along with
identifications of the strong lines.  The telluric identifications
($\earth$) refer to residuals left over from the removal of the
atmospheric H$_2$O and O$_2$ bands.  \label{fig_id_opt}}
\end{center}
\end{figure}

  The optical spectrum of the M33 nucleus is shown in
Figure~\ref{fig_id_opt} with the strong lines identified
(\cite{bic86}, 1987).  Our spectrum is similar to that given in
Figure~1 of Schmidt et al.\ (1990) except ours has a slightly steeper
slope and a more pronounced Paschen jump.  In addition, our spectrum
is absolutely calibrated while the Schmidt et al.\ (1990) spectrum was
only relatively calibrated.

  The near-IR observations were made using the Steward Observatory
infrared spectrometer, FSpec (\cite{wil93}), on the Multiple Mirror
Telescope (MMT) on 1996 June 21, 22 (UT).  The spectra
were taken using the medium resolution, 300 lines/mm grating.  Four
wavelength settings were used, $J$-band from 1.15 to 1.35~$\mu$m,
$H$-band from 1.54 to 1.77 $\mu$m and two settings at $K$-band, from
2.015 to 2.23 $\mu$m and from 2.19 to 2.41 $\mu$m.  The spectral
resolution, as measured via OH sky emission lines, or Neon Krypton
calibration lamps in the case of the long wavelength K-band,
corresponded to approximately 2.5 resolution pixels (21 \AA), or R
$\approx$ 590 at 1.25~$\mu$m, R $\approx$ 780 at 1.65 $\mu$m, R
$\approx$ 1000 at 2.12 $\mu$m and R $\approx$ 1100 at 2.3 $\mu$m.  The
same observing procedure was used on both nights and with all grating
positions.  The spectrometer has a slit size of $1.2\arcsec \times
32\arcsec$ on the MMT.  This allowed for four unique positions as the
M33 nucleus was stepped $\approx 6\arcsec$ along the slit.  The
integration time at each position was 2 minutes giving a total of 8
minutes per set of observations. A total integration of 24 minutes was
obtained at both $J$- and $H$-band.  At 2.12 $\mu$m (short
$K$-band) and 2.30 $\mu$m (long $K$-band), the total integration times
were 40 and 32 minutes respectively.

  The reductions made use of a dark-current-corrected flat field
obtained from a uniformly illuminated screen. After dark current had
been subtracted and the flat field divided into the raw
two-dimensional images, sky emission (and additional thermal
background in the K-band) was removed by subtracting one slit position
from the next. In many cases, a few percent scaling was required to
get the telluric atmospheric OH features to disappear entirely.
Background renormalization by a few percent was also required to
remove fluctuations in thermal background between integrations in the
K-band.  Interspersed between our M33 observations, we obtained
spectra at a similar airmass of HR 410, a late F dwarf, to be used to
remove telluric artifacts. Intrinsic features in the spectrum of this
star were removed using a Kurucz model atmosphere, with T$_e$ = 6500,
log(g) = 4.5, solar abundance. Dividing the spectrum of M33 by this
modified stellar spectrum allowed us to cancel features due to the
atmosphere.  While our near-infrared spectra are not flux calibrated
(due to slit loss and non-photometric conditions) we can determine the
accurate relative flux distributions for each grating position.

  The long slit of FSpec allowed us to look for evidence of light
extended beyond the nuclear region.  While there likely is some low
level flux outside the nucleus, dividing one slit position by another
in the reductions revealed no structure outside of the nuclear point
source.  Measurement of the full-width, half-maximum flux of the
nucleus along the slit also reveals the nucleus to be unresolved
spatially.  Typical seeing during the observations was very good,
approximately 1 to $1.5\arcsec$; however by the end of the
integration, the multiple mirrors of the MMT drifted slightly relative
to each other on the sky, reducing the effective seeing to $\sim$
$2\arcsec$.

\begin{deluxetable}{lccc}
\footnotesize
\tablewidth{0pt}
\tablecaption{Near-Infrared Line Identifications and Equivalent Widths
\label{tab_nir}}
\tablehead{\colhead{Ion} & \colhead{Transition} &
           \colhead{Vacuum $\lambda$} & \colhead{Equiv.\ Width} \\ 
           \colhead{} & \colhead{} & \colhead{[$\mu$m]} & 
             \colhead{[\AA]} }
\startdata
\multicolumn{4}{c}{J band} \nl \hline
??          &             &      1.196         &   1.5   \nl
\ion{H}{1}  & (3-5)       &      1.2818        &   1.5   \nl \hline
\multicolumn{4}{c}{H band} \nl \hline
$^{12}$CO   & (3,0)       &      1.558         &   1.3     \nl
??          &             &      1.565         &   0.8     \nl
\ion{Mg}{1} & (4s-4p, tr) &1.577, 1.575, 1.575 & $\sim3.0$\tablenotemark{a} \nl
$^{12}$CO   & (4,1)       &      1.578         &   $''$  \nl
\ion{Si}{1} &             &      1.5894        &   2.5     \nl
\ion{Si}{1} & (4p-5s)     &      1.596         &   0.3     \nl
$^{12}$CO   & (5,2)       &      1.598         &   1.5     \nl
\ion{H}{1}  & (4-13)      &      1.6109        & $\le$1    \nl
$^{12}$CO   & (6,3)       &      1.619         &   2.5      \nl
\ion{H}{1}  & (4-12)      &      1.6407        &   1.8     \nl
??          &             &      1.6525        &   1.3     \nl
$^{12}$CO   & (8,5)       &      1.662         &   4.0     \nl
\ion{H}{1}  & (4-11)      &      1.6811        &  $<$1      \nl
\ion{Mg}{1} & (4s-4p)     &      1.7146        &   4.6      \nl 
$^{12}$CO   & (10,7)      &      1.7067        &   2.0     \nl
\ion{Fe}{1} &             &      1.7307        &   1.2     \nl
\ion{Si}{1} &             &      1.7332        &   2.0\tablenotemark{a} \nl
\ion{H}{1}  & (4-10)      &      1.7362        &   $''$   \nl  \hline
\multicolumn{4}{c}{K band} \nl \hline
??          &             &      2.0443        &   1.0      \nl
\ion{Fe}{1} &             &      2.0704        & $\sim1$    \nl
\ion{Si}{1} & (4p-5s)     &      2.1360        &  $\le$1     \nl
\ion{H}{1}  & (4-7)       &      2.1661        &   2.0       \nl
\ion{Ti}{1} &             &      2.1789        & $\sim$0.3   \nl
\ion{Si}{1} &             &      2.1885        &   1.0       \nl
\ion{Na}{1} &             &      2.2062        &   $\sim$1   \nl
\ion{Na}{1} &             &      2.2090        &   $<$0.5    \nl
$^{12}$CO   &  (2,0)      &      2.2935        & $\sim$10    \nl
$^{12}$CO   &  (3,1)      &      2.3227        & $\sim$8     \nl
$^{13}$CO   &  (2,0)      &      2.3448        & $\sim$3.5   \nl
$^{12}$CO   &  (4,2)      &      2.3525        & $\sim$9     \nl
$^{13}$CO   &  (3,1)      &      2.3739        & $\sim$4     \nl
$^{12}$CO   &  (5,3)      &      2.3830        & $\sim$5.5   \nl
$^{13}$CO   &  (4,2)      &      2.4037        & $\sim$2.5   \nl
\enddata
\tablenotetext{a}{Blended}
\end{deluxetable}

\section{The Starburst Model}

  To model the spectral energy distribution (SED) of a starburst
region, we need to model the stellar and gas emission accurately,
including the modification of this emission by absorption and
scattering by dust in the starburst region.  Other input parameters
include the distance to the starburst, the radial velocity of the
starburst, and the dust column in our Galaxy towards the starburst.
For the M33 nucleus, the distance is 795 kpc, the foreground dust
column has an $E(B-V) = 0.07$ and was assumed to be Milky Way (MW)
dust with $R_V 
= 3.1$ (\cite{car89}), and the radial velocity is -172~km~s$^{-1}$
(\cite{zar89}; \cite{van91}).

  The starburst model is an improved version of that used by Gordon et
al.\ (1997).  The stars and gas in a starburst are modeled using
Leitherer et al.\ (1995, 1998; hereafter L98) and Fioc \&
Rocca-Volmerage (1997, 1998; hereafter PEGASE) stellar evolutionary
synthesis (SES) models.  We used two SES models in an attempt to
estimate the uncertainties in the model SEDs due to uncertainties in
the assumed stellar evolutionary tracks, stellar spectra, and
computational algorithms.  Both models were run for burst and constant
star formation scenarios, an initial mass function (IMF) with a
Salpeter 
(1955) slope ($\alpha = 2.35$), a mass range of 0.1--100 M$_{\sun}$,
and solar metallicity.  The L98 model used Geneva evolutionary tracks
and the PEGASE model used Padova evolutionary tracks.  

  The choice of a Salpeter IMF for stellar masses greater than $\sim
0.85~M_{\sun}$ is supported by the observations in Local Group
clusters and associations (\cite{hun97}). Furthermore, IMF8, found by
Rieke et al.\ (1993) to be optimum for the M82 starburst, has a net
high mass slope similar to the Salpeter value. There are no direct
constraints on the form of the IMF below $0.85~M_{\sun}$ in
starburst-type regions, but the local IMF has a slope of $\alpha \sim$
1 below masses of $\sim$$0.7~M_{\sun}$ (e.g., \cite{sca98}).  For this 
work, we assume the IMF has the form
\begin{equation}
\label{eq_imf}
N(m) = Cm^{-\alpha} = \left\{
   \begin{array}{ll}
   Cm^{-2.35} & m > 0.7~M_{\sun} \\
   1.619Cm^{-1} & m < 0.7~M_{\sun} \\
   \end{array} \right.
\end{equation}
where $C$ is the normalization constant of the IMF.  The use of a
different IMF below $0.85~M_{\sun}$ changes the total mass needed for
the burst, but does not affect the computed SED for the ages
appropriate for the M33 nucleus.  For example, using
the IMF given in equation~\ref{eq_imf} would imply 59\% of the mass
given by the assumption of a Salpeter slope for the entire IMF. To
obtain realistic masses, we have therefore multiplied the model values
by 0.59.

  The metallicity of the M33 nucleus should be similar to the
surrounding region and the central abundances of M33 are approximately
solar (\cite{gar97}; \cite{mon97}).  SEDs were produced for ages up to
$1 \times 10^{9}$ years for L98 and $1.9 \times 10^{10}$ years for
PEGASE.  Both models used the Lejeune, Cuisinier, \& Buser (1997)
spectral library which has a resolution of $\sim$40~\AA.  An
additional input to both models is the fraction of Lyman continuum
photons which contribute to the nebular emission (the rest are assumed
to be absorbed by dust).  Since the nucleus of M33 does not show any
emission at H$\alpha$, the contribution to the SED from
nebular emission is negligible and the value of this model parameter
is unimportant.

  Both SES models produce similar SEDs for ages between 10 and 100
million years (Myrs).  For ages $<$ 10 Myrs, there are quite
significant differences in the optical/IR between the SEDs which can
be traced to the different evolutionary tracks used (Geneva versus
Padova; \cite{fio97}).  After 100 Myrs, the PEGASE model produces
consistently higher IR fluxes, which is due to the inclusion of the
thermally pulsating asymptotic giant branch (TP-AGB) evolutionary
tracks in this model (\cite{fio97}). 
These tracks are not included in the L98 model, limiting its validity
to ages $< 100$ Myrs (C.\ Leitherer 1998, private communication). 

  The effects of dust were modeled using a Monte Carlo radiative
transfer code (\cite{wit77}; \cite{wit96}; \cite{gor97};
\cite{wit98}).  Calculation of the radiative transfer through dust is
crucial when any geometry other than the screen geometry is suspected.
Our radiative transfer model was run for dust global geometries
characterized as dusty (dust, stars, and gas uniformly mixed together)
and shell (a sphere of stars and gas enshrouded by a shell of dust).
These two geometries define the extremes one would expect for star
formation regions.  The local dust geometry was characterized with
inter-clump to clump density ($k_2/k_1$) ratios between 1.0 and 0.001
and a clump filling factor of 15\%.  In a given model, the physical
properties of the dust grains ($\tau_{\lambda}$, albedo, and
scattering phase function) were characterized by either MW ($R_V =
3.1$, \cite{car89}) or Small Magellanic Cloud Bar (\cite{gor98}) type
dust (\cite{gor97}).  The amount of dust is parameterized by the V
band optical depth, $\tau_V$.  The dust models were run for $\tau_V$
values ranging from 0.25 to 50.

  Combining the SES and dust radiative transfer models provides a
model for the SEDs of starburst regions which accurately includes
stars, gas, {\em and} dust.  The parameter space covered by the
starburst model is large and, as a result, an algorithm to determine
which of the model SEDs fit the observed SED was needed.  Following
Gordon et al.\ (1997), we use the $\tilde{\chi}^2$ (reduced $\chi^2$)
method.  The $\tilde{\chi}^2$ was computed using
\begin{equation} 
\label{eq_chisqr}
\tilde{\chi}^2 = \frac{1}{d} \sum_{i=1}^{n} \left[
    \frac{F_o(\lambda_i) - F_m(\lambda_i)}{\sigma(\lambda_i)}
    \right]^2
\end{equation}
where $n$ is the number of points in the observed SED, $d$ is the
number of degrees of freedom, $F_o(\lambda_i)$ is the observed flux at
the $i$th point in the SED, $F_m(\lambda_i)$ is the same for the model
flux, and $\sigma(\lambda_i)$ is the $i$th uncertainty in the observed
flux (\cite{tay82}).  In the paper by Gordon et al.\ (1997), a
typographical error resulted in the square being left off the term in the sum
of Equation~\ref{eq_chisqr}.  In an attempt to also account for the
observational 
uncertainties in the SES and radiative transfer model predictions, we
have added, in quadrature, a 15\% (10\%) uncertainty to $\sigma(\lambda_i)$
of the ultraviolet/infrared (optical) observations.  The uncertainties
associated with the SES models 
are due to their use of observational spectra to calibrate the stellar
SEDs (\cite{lej97}).  The uncertainties associated with the radiative
transfer model are due to the use of observationally determined values
of the physical properties of dust grains.  The 10-15\% uncertainty
quoted above for the starburst model is a reasonable assessment of these
uncertainties.  In fitting the model SEDs to the observed SED, there are
4 parameters which are derived from the observed SED (age, mass,
$\tau_V$, and $k_2/k_1$).  Thus, $d = 17 - 4 = 13$.  In our fitting
algorithm, all of the model SEDs were computed for the input distance,
reddened by the foreground dust, and then shifted to the rest frame of
M33.

\section{Model Fits \label{sec_fits}} 

\begin{figure}[tbp]
\begin{center}
\plotone{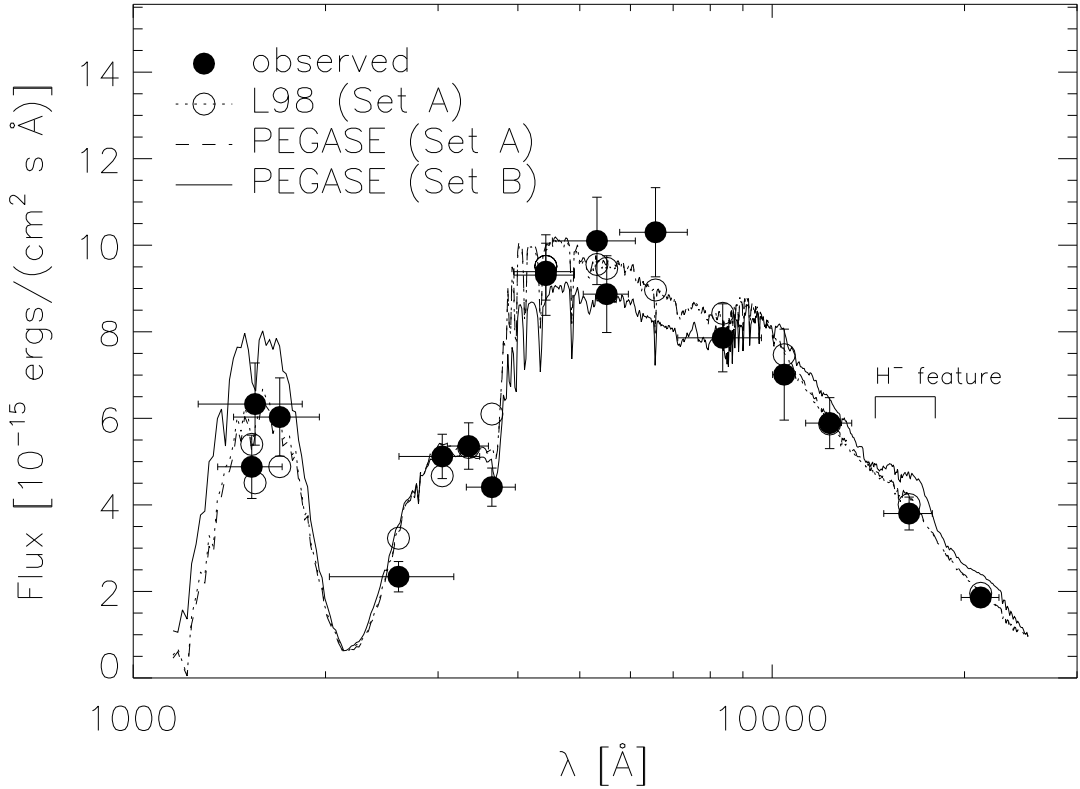}
\caption{The observed photometric SED and best fit model SEDs for both
Set A and B are plotted.  The horizontal error bars on the observed SED points
represent the bandwidth of the observations.  The open circles give
the fluxes in the bands corresponding to the observed photometry.
They were determined using the band response curves and the L98 best
fit model SED.  The $1.6~\micron$ H$^{-}$ feature is identified (see
\S\ref{sec_fits}). \label{fig_obs_sed}}
\end{center}
\end{figure}

  The model SEDs were fitted to the photometric data given in
Table~\ref{tab_obs_sed}.  The model fluxes in the observed photometric
bands, $F_m(\lambda_i)$, were calculated by convolving the band
transmission curves (\cite{bes88}; \cite{bes90}; \cite{bir96};
\cite{ste97}; M.\ Bessell 1998, private communication) with the full
resolution model SEDs.  We examined all model SEDs which had
$\tilde{\chi}^2 < 2.14$ as this includes all model SEDs which fit with
a $>$1\% significance, i.e.\ the probability of the correct model
getting a higher $\tilde{\chi}^2$ [$P_{13}(\tilde{\chi}^2 \geq 2.14)$]
is only 1\%.  These model SEDs were grouped into two sets, (A) those
with $\sim$70 Myr old burst of star formation and (B) those with
$\sim$1 Gyr of constant star formation.  For both sets, no models
with a dusty geometry fit the observations.  The best fit model
parameters for set A and B are given in Table~\ref{tab_best_fit}.  The
uncertainties on these parameters were determine by the 99\%
confidence level as described above.  The observed photometric SED of
the M33 
nucleus is plotted in Figure~\ref{fig_obs_sed} along with the best fit
model SEDs in from both Set A and B.  The Set A models fit the
photometric SED better than the Set B models according to the
$\tilde{\chi}^2$ values (Table~\ref{tab_best_fit}).  This can be seen
in Figure~\ref{fig_obs_sed} where the best fit Set B model SED is
consistently too high in the far-UV, too low in the optical, and too
high in the IR.

\begin{deluxetable}{lccc}
\tablewidth{0pt}
\tablecaption{Best Fit Model Parameters\tablenotemark{a} \label{tab_best_fit}}
\tablehead{ & \colhead{L98} & \multicolumn{2}{c}{PEGASE} \\
           \colhead{parameter} & \colhead{Set A} & \colhead{Set A} & 
           \colhead{Set B} }
\startdata
star formation type & burst & burst & constant \nl
geometry & shell & shell & shell \nl
dust type & Milky Way & Milky Way & Milky Way \nl
age [years] & $65^{+30}_{-30} \times 10^6$ & 
   $70^{+110}_{-30} \times 10^6$ & 
   $1^{+0.2}_{-0.2} \times 10^9$ \nl
mass [$M_{\sun}$] & $6.4^{+3.7}_{-2.7} \times 10^{5}$ & 
   $6.9^{+2.3}_{-2.2} \times 10^{5}$ & 
   $1.2^{+0.2}_{-0.1} \times 10^{6}$ \nl
$k_2/k_1$ & $0.5^{+0.5}_{-0.45}$ & $0.5^{+0.5}_{-0.45}$ & $1.0^{+0}_{-0.5}$ \nl
$\tau_V$ & $2^{+1.25}_{-0.25}$ & $2.13^{+0.75}_{-0.75}$ & 
   $2^{+0.12}_{-0}$ \nl
$\tau_{V,{\rm eff}}$ & $1.26^{+0.36}_{-0.18}$ & $1.33^{+0.27}_{-0.5}$ & 
   $1.26^{+0.08}_{-0}$ \nl
$M/L_V$ [$M_{\sun}/L_{V,\sun}$] & $0.13^{+0.03}_{-0.06}$ & 
   $0.12^{+0.1}_{-0.05}$ & $0.22^{+0.03}_{-0.03}$ \nl
$\tilde{\chi}^2$ & $1.39$ & $1.44$ & $1.96$ \nl
\enddata
\tablenotetext{a}{The uncertainties quoted are for the 99\% confidence
level ($\sim$$3\sigma$).}
\end{deluxetable}

\begin{figure}[tbp]
\begin{center}
\plotone{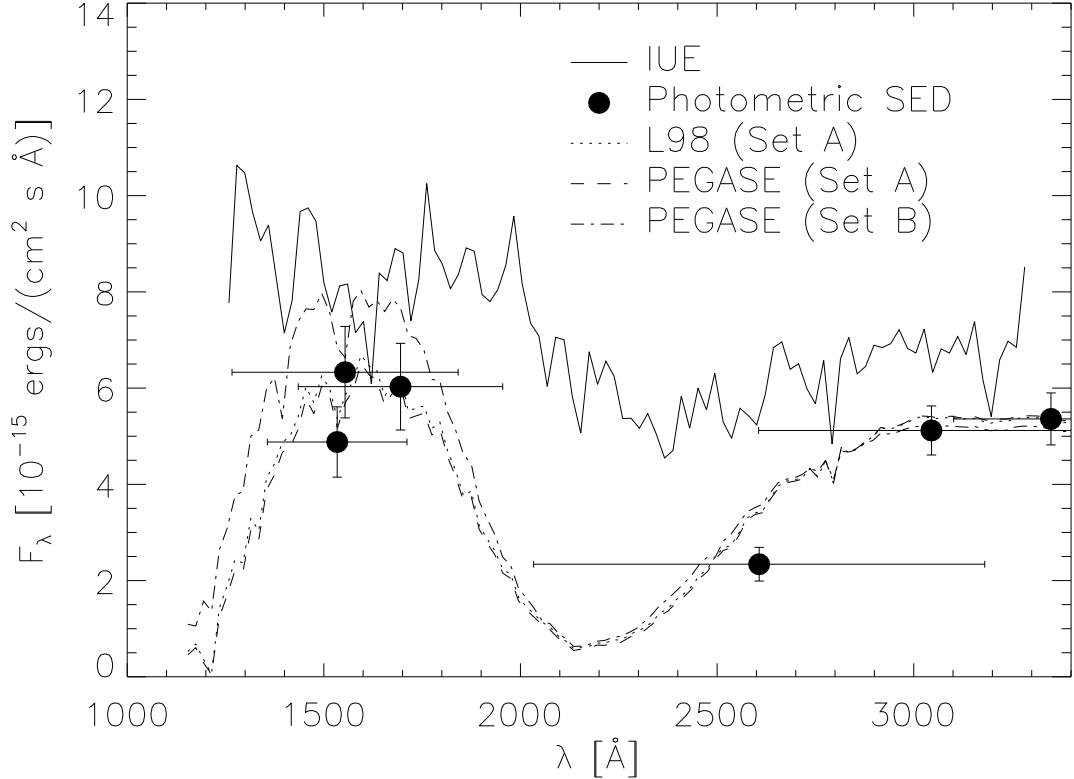}
\caption{The IUE spectrum of the M33 nucleus and surrounding region is
plotted.  In addition, the photometric SED and best fit model SEDs for
Sets A and B are plotted.
\label{fig_iue}}
\end{center}
\end{figure}

\begin{figure}[tbp]
\begin{center}
\plotone{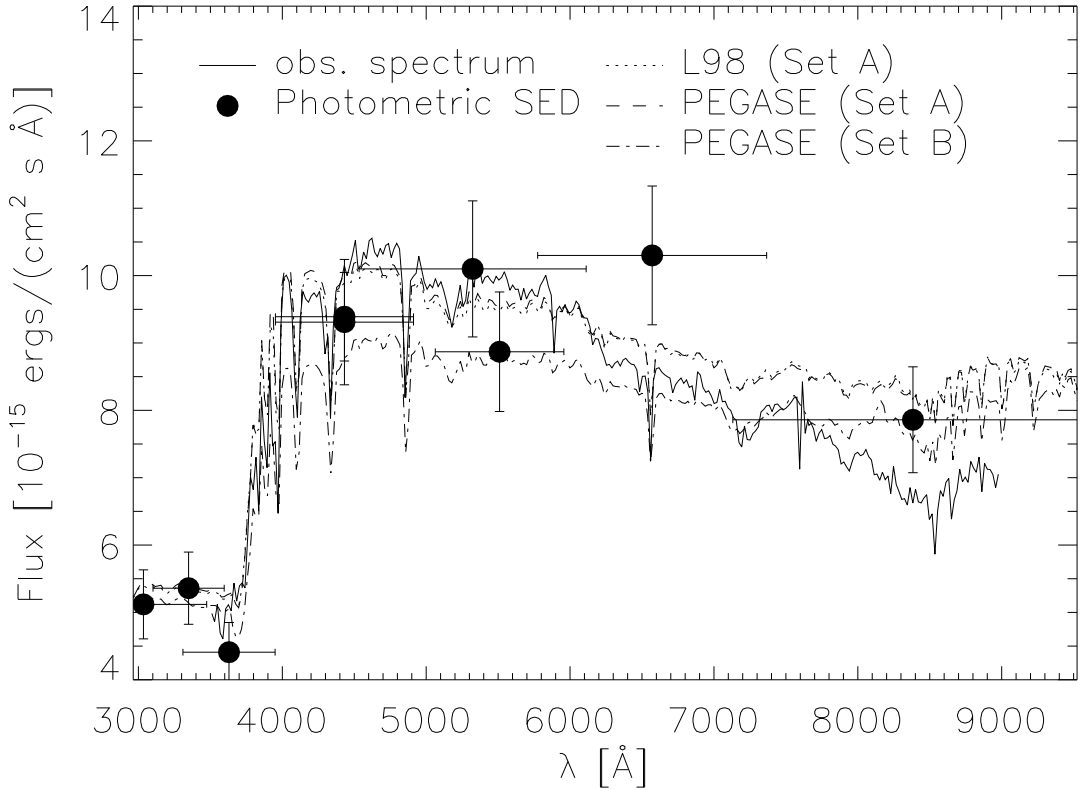}
\caption{The optical spectrum is plotted along with the best fit model
SEDs from Set A and B.  The optical spectrum has been rebinned to
match the resolution of the model SEDs.
\label{fig_optical}}
\end{center}
\end{figure}

\begin{figure}[tbp]
\begin{center}
\plottwo{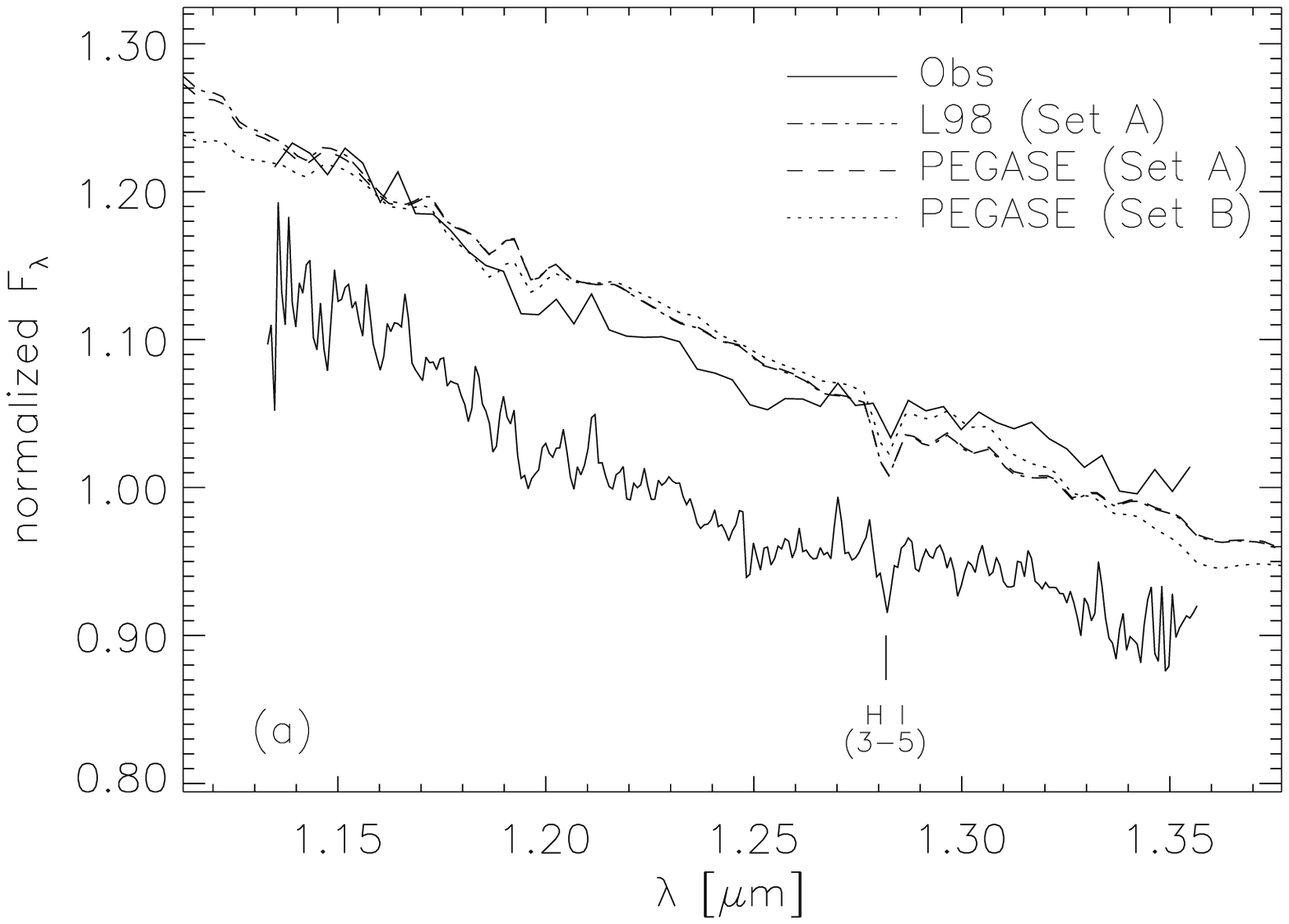}{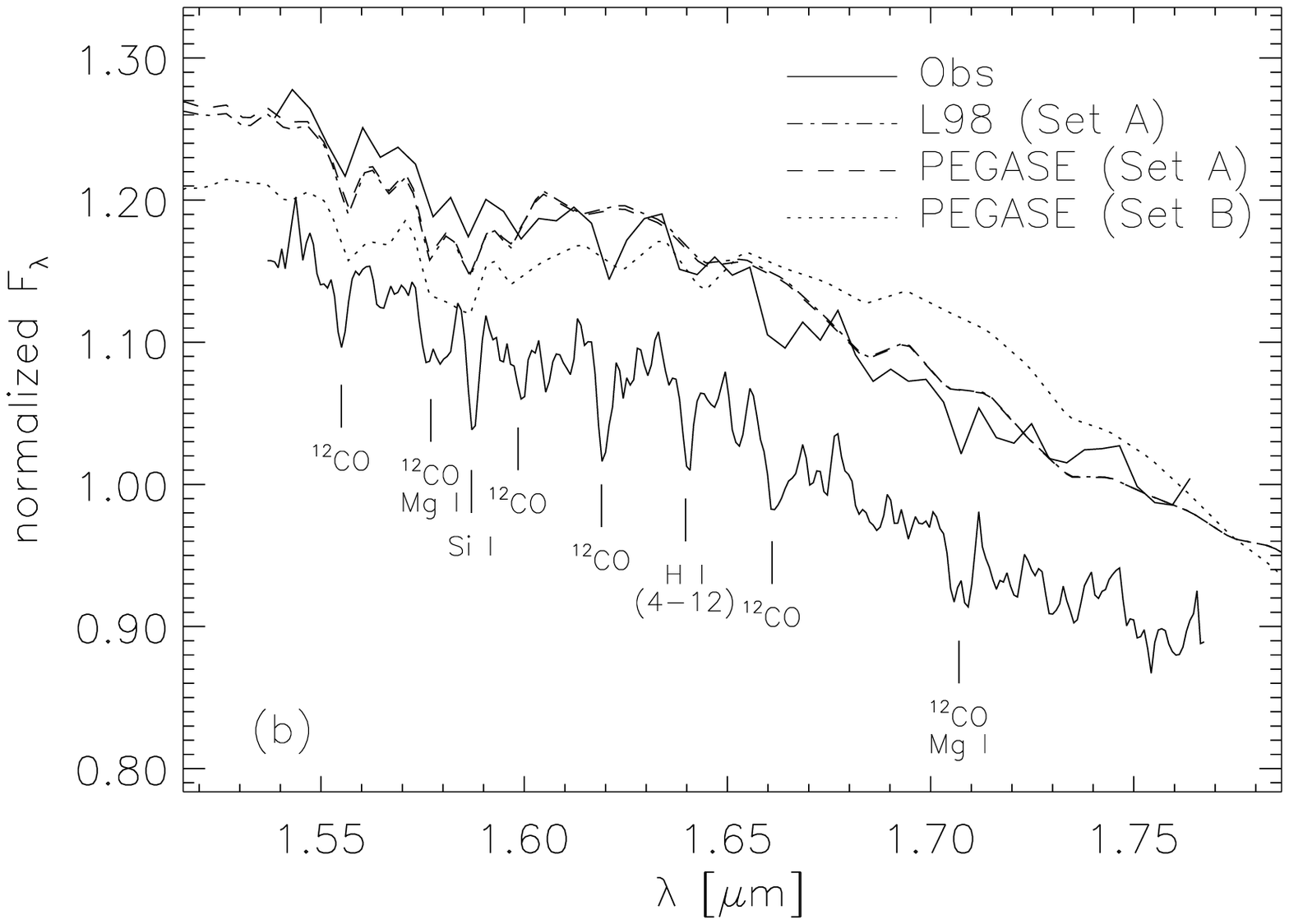} \\
\plottwo{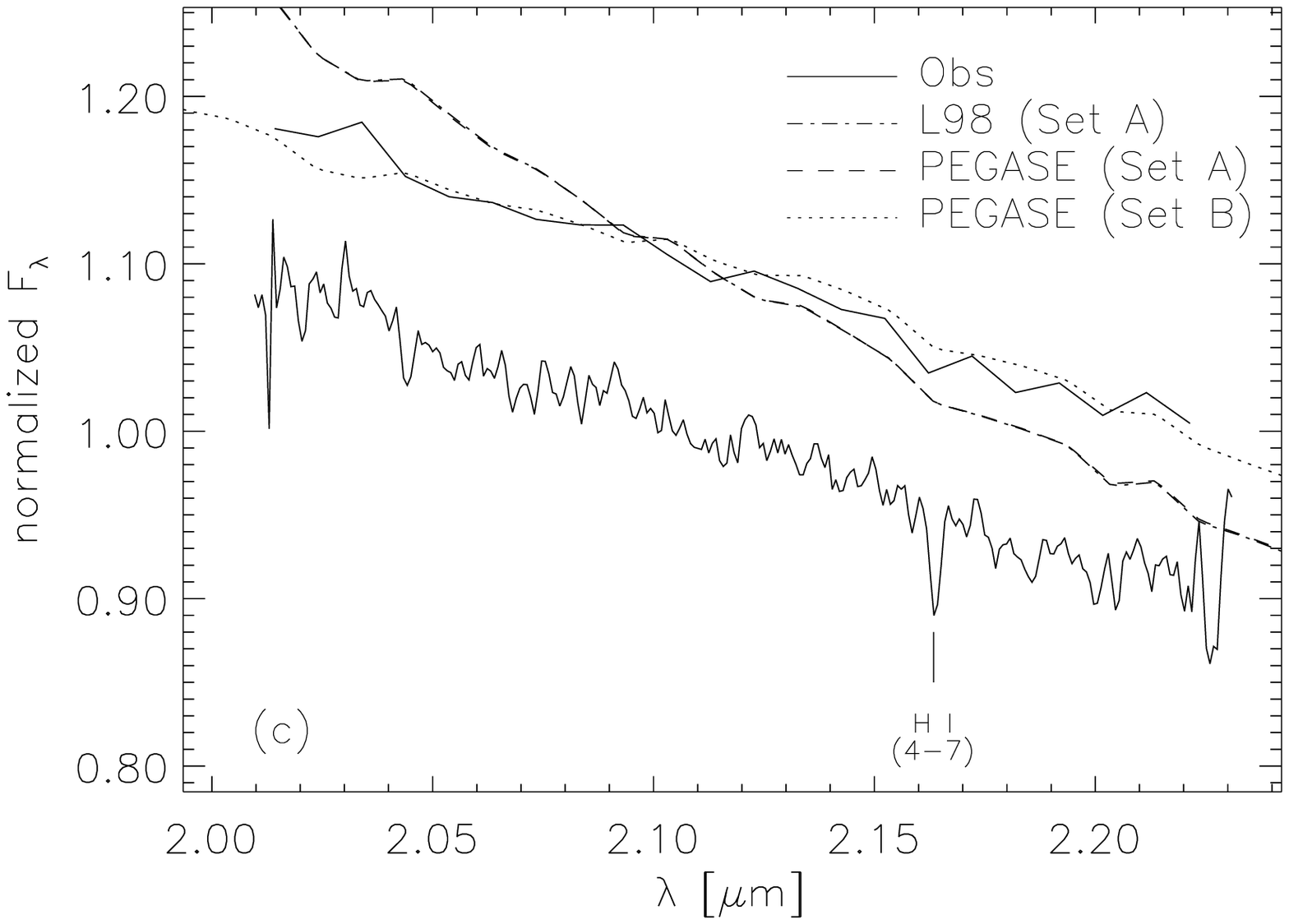}{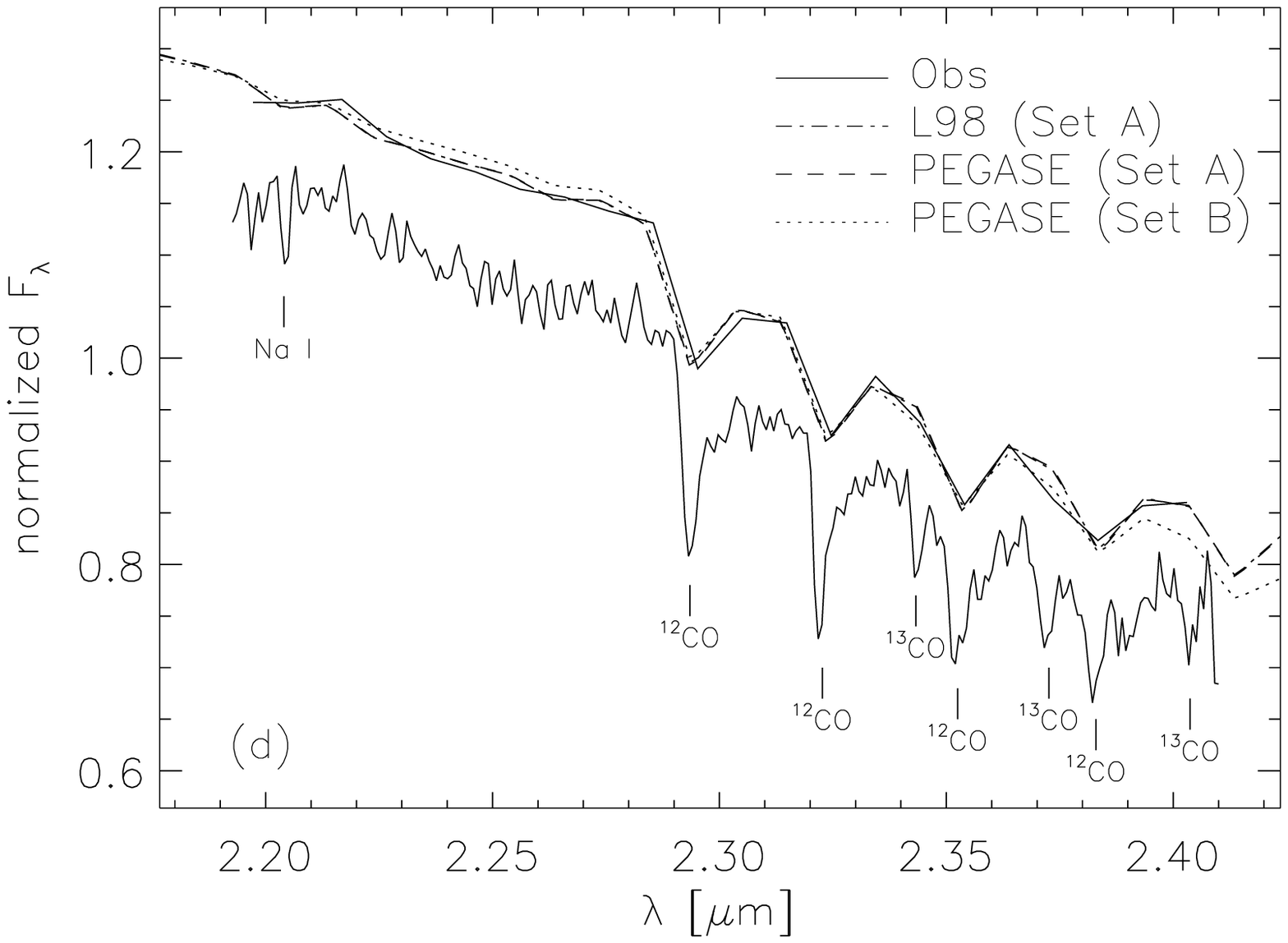}
\caption{The observed IR spectra are plotted along with the best
fit model SEDs.  The J band spectrum is plotted in (a), the H band in
(b), and the two K band spectra in (c) and (d).  As the observed
spectra are only relatively calibrated, the model SEDs are plotted
normalized to the average value of the observed spectrum.  The lower
observed spectrum is plotted at the observed resolution and the upper
observed spectrum is plotted at the model resolution (100 \AA\ for J
\& H band, 200~\AA\ for the K band).
\label{fig_ir_spectra}}
\end{center}
\end{figure}

  The validity of Sets A and B can be tested further with the
spectroscopic observations. The UV, optical, and IR spectra of the M33
nucleus are plotted in Figures~\ref{fig_iue}-\ref{fig_ir_spectra}.  As
the spectroscopic data were taken with different apertures, it is not
possible to fit simultaneously the UV to near-IR spectroscopy of the
M33 nucleus.  In addition, the use of relatively large apertures
results in the flux from the nucleus being diluted by flux from the
surrounding region.  Therefore, we have used the photometric data
along with the spectra for the comparisons with the best fit model
SEDs. 

  Figure~\ref{fig_iue} plots the IUE spectrum along with the
photometric SED and the best fit model SEDs from Sets A and B.  Due to
the faintness of the M33 nucleus, the S/N in the IUE spectrum is quite
low.  We display the spectrum with 20~\AA\ bins to increase the S/N
and match the resolution of the model SED.  The IUE spectrum obviously
includes flux from both the nucleus and the surrounding region since
the photometric SED is always below the IUE spectrum.  The shape of
the IUE spectrum is flatter than the photometric and model SEDs,
implying that the UV spectrum of the region surrounding the nucleus
has a much smaller depression at 2175~\AA\ and, therefore, is
attenuated by less dust than the nucleus.  This behavior arises from
mixing flux from regions which have differing amounts of dust: the
least attenuated region will contribute disproportionately to the
flux, washing out the signatures of dust.

  The optical spectrum of the M33 nucleus is plotted in
Figure~\ref{fig_optical} along with the best fit model SEDs from Sets A 
and B.  Unlike the IUE spectrum, the optical spectrum is roughly
consistent with the photometric SED.  While the steeper slope makes
the observed 
spectrum slightly bluer than all of the model SEDS, the strengths
of the features are comparable, with some notable exceptions.  For
example, the observed strength of the \ion{Na}{1} $\lambda$5889
(1.8~\AA) can be traced to the presence of a large amount of
interstellar material.  This is consistent with our detection of a
shell of dust with $\tau_V \sim 2$ enshrouding the nucleus.  Both the
differing line strengths and the bluer color could arise from
inclusion of flux from the region surrounding the nucleus as the
observed spectrum includes light from a $3\arcsec \times 10\arcsec$
region.  The 
general agreement with the nuclear photometric levels would still be
consistent if some of the nuclear flux spilled out of the slit (the
seeing was $\sim$$2\arcsec$).

  The optical spectrum shows no significant emission in H$\alpha$ or
any other line.  Set B models predict H$\alpha$ in emission with an
equivalent width of $\sim$100~\AA\ which would easily have been seen
in the observed spectrum.  The SES models predict the strength of H$\alpha$,
but do not add them to their output SEDs.  Thus, no emission lines are
seen in Figure~\ref{fig_optical} for the Set B model SED even through
they would be very prominent.  In
comparison, the 
Set A models predict no detectable H$\alpha$ emission, in good
agreement with the observations.

  The infrared spectra are plotted in Figure~\ref{fig_ir_spectra}.
All the models do a fairly good job of fitting them, as expected since
all the model SEDs fit the photometric SED.  Most of the observed
absorption lines are present in the model SEDs, but not necessarily at
the same strength.  The model SEDs have much weaker Brackett lines
than observed (\ion{H}{1} (4-12) and \ion{H}{1} (4-7) are good
examples, but we also detect \ion{H}{1} (4-10)), and they do not show
the $^{13}$CO lines. Furthermore, the $^{12}$CO bands are somewhat
weak in the models, both at the first overtone near $2.3~\micron$ and the
second near $1.6~\micron$. Since the inputs to the infrared models have
greater uncertainty and are less complete than those for the optical,
it is not surprising that this is also the region where the models are
most deviant.

  There is only one infrared spectral feature that differs
significantly between the Set A and B models, the broad $1.6~\micron$
H$^{-}$ peak is seen in stars with $T_{\rm eff} < 6000 K$ (\cite{lan92}).
This feature is not seen in the observations (Figure~\ref{fig_obs_sed}
and \ref{fig_ir_spectra}b), which suggests that the Set A models
better describe the M33 nucleus than the Set B models.  The Set B
models reproduce the slope of the observed short K band spectrum
(Figure~\ref{fig_ir_spectra}c), but given the difficulties in
determining general slopes in the data, this disagreement is probably
not significant.  

    Because of its sensitivity to the stellar luminosity, the CO band
strength has proven useful in determining the age of starbursts. This
parameter is commonly expressed as a "CO index" based on the average
depth of the feature in intermediate width photometric bands
(\cite{fro78}).  We 
have used the formulation of Doyon, Joseph, \& Wright (1994) to
compute CO indices 
for our observations of M33 and for the L98 and PEGASE synthetic
spectra. For M33, the index is approximately 0.21. The synthetic
spectra show obvious shortcomings in both cases, such as discontinuous
changes in index strength for the PEGASE models and an index that
never reaches the highest observed strengths for L98. For the
starburst models of \cite{rie93}, the CO index of the
individual stellar types was carefully calibrated empirically against
the extensive available photometric observations. The run of CO index
with starburst age for these models (\cite{shi96}) does
not have the problems we find in the synthetic spectra, and we prefer
it for comparison with M33. There are two cases that represent a
relatively short burst, either a star formation rate Gaussian in time
with a full width at half maximum of 5 Myr, or a burst that starts
abruptly and decays with an exponential time constant of 20 Myr. These
cases produce an index of 0.21 after 50 to 70 Myr, in excellent
agreement with the parameters of the burst we have derived from the
spectral synthesis in the optical and ultraviolet.

\section{Discussion}

\subsection{Nuclear Mass}

  Assuming the rotation curve of M33 is flat within $4\arcsec$,
Kormendy \& McClure (1993) calculate an upper limit of $2.1 \times
10^{6} ~M_{\sun}$ on the mass of the nucleus.  Lauer et al.\ (1998)
calculate a lower limit by determining a core mass of $2 \times 10^{4}
~M_{\sun}$.  A better estimate of the mass in the starburst region
should be possible from the measured velocity dispersion (21 km
s$^{-1}$, Kormendy \& McClure 1990) and the size of the starburst.

  We measured the FWHM of the PC images of the M33 nucleus by fitting
them with 2-dimensional Gaussian convolved with the TinyTim point
spread function (PSF) appropriate for each filter (\cite{kri97}).  It
was not possible to fit the nucleus in the F160BW filter due to its
low signal.  The results are given in Table~\ref{tab_nuc_size}.  Lauer
et al.\ (1998) found the FWHM was $0\farcs 07$ (0.27 pc) in the F555W
image using a different analytic form for the nuclear profile and
taking better care to correct for the effects of the PSF and aliasing.
The increasing size of the nucleus as a function of wavelength
confirms the color gradient measured by Kormendy \& McClure (1993) and
Lauer et al.\ (1998).

\begin{deluxetable}{lccc}
\tablewidth{0pt}
\tablecaption{Nucleus Size \label{tab_nuc_size}}
\tablehead{ & \multicolumn{3}{c}{FWHM} \\ 
           \colhead{Filter} & \colhead{[pixels]} & 
           \colhead{[$\arcsec$]} & \colhead{[pc]} }
\startdata
F300W  & 2.65 & 0.12 & 0.46 \nl
F555W  & 4.33 & 0.20 & 0.77 \nl
F814W  & 4.62 & 0.21 & 0.81 \nl
F1042M & 5.58 & 0.25 & 0.98 \nl
\enddata
\end{deluxetable}

  Because they dominate the light, the observed velocity dispersion
and size reflect only properties of the more massive stars.  This is
confirmed by the composite spectral type of the nucleus, for which a
range from late-A to early-F main sequence has been reported
(\cite{mor57}; \cite{van91}; \cite{mas96}; B. Garrison, private
communication).  Assuming equipartition in kinetic energies, the true
velocity dispersion ($\sigma_t$) of the nucleus is then
\begin{equation}
\sigma_t = \sigma_o \left( \frac{m_o}{m_t} \right)^{1/2}
\end{equation}
where $\sigma_o$ is the observed velocity dispersion, $m_t$ is the
average mass of the stars in the nucleus, and $m_o$ is the average
mass of the stars contributing to $\sigma_o$.  Assuming the IMF given
in equation~\ref{eq_imf}, $m_t = 0.84~M_{\sun}$.  From the
spectral type of the nucleus, $m_o \approx 3 ~M_{\sun}$.  Thus, the
true velocity dispersion of the nucleus is 40 km s$^{-1}$. The virial
mass is computed from
\begin{equation}
M_{VT} = \frac{3R\sigma_t^2}{G} = 698R\sigma_t^2
\end{equation}
where R is the effective radius of the nucleus in pc, $\sigma_t$ is in
km s$^{-1}$, and $M_{VT}$ is in $~M_{\sun}$.  Estimating $R$ for
the nucleus is nontrivial and so we will use a measured $R$ value and
compute a lower limit.  We assume $R$ to be the size of the nucleus as
measured in the F814W image (HWHM $= R = 0.41$ pc) since $\sigma_o$
was measured from the \ion{Ca}{2} infrared triplet (\cite{kor93}).
The mass of the nucleus is then $> 4.6 \times 10^{5}~M_{\sun}$.  This
is similar to the $6-7 \times 10^{5}~M_{\sun}$ of the best fit Set A
models.  

  In addition to the starburst population, it is likely there is at
least one other stellar population in the nuclear region.  The
presence of an older ($\sim$1 Gyr) underlying stellar population with
a large $M/L_V$ is strongly implied by the growth of surface
brightness and stellar counts approaching the nuclear region
(\cite{min93}) and the mass ($2.1 \times 10^6 M_{\sun}$) of the inner
$4\arcsec$ of M33 (\cite{kor93}).  Within 4$\arcsec$, 3/4 of the
observed V band flux comes from the nucleus (\cite{kor93}).  This
implies that the other stellar population in the nuclear region has a
$M/L_V$ ratio of 5.5 assuming the Set A best fits or 1.8 assuming the
Set B best fit.  Thus, the mass required by the Set A models ($6-7
\times 10^5 M_{\sun}$) is more consistant with the available data on
the nuclear region.

\subsection{Stellar Population \label{sec_stell_pop}}

  The photometric and spectroscopic evidence indicate that Set A 
models fit the observations of the M33 nucleus better than the Set B
models, as does  
the mass determination.  The Set A models are consistent
with the small size of the nucleus and flat profile of the inner disk
of M33.  The flat profile implies there is a lack of
a substantial density gradient which could funnel gas to the nucleus
(\cite{kor93}), which means fueling continuous star formation (as Set
B fits would require) would be difficult.  The stellar population of
the M33 nucleus is described by a 35 to 180 Myr burst of star
formation with a mass of 0.4-1.1~$\times 10^{6}~M_{\sun}$
(Table~\ref{tab_best_fit}).  The best fit SEDs in Set A for both SES
models give very similar results.  They imply the best description of
the M33 nucleus is a burst of star formation 70-75 Myrs old and a mass
of $\sim$0.7~$\times 10^{6}~M_{\sun}$ (Table~\ref{tab_best_fit}).

  Our finding that the M33 nucleus is well described by a single burst
of star formation enshrouded by a significant amount of dust is quite
different than previous studies (\cite{oco83}; \cite{cia84};
\cite{sch90}).  In these studies, the role of dust was underestimated
and, as a result, luminous young and old stellar populations were both
needed to reproduce the relatively red (flat) spectrum of the M33
nucleus. For example, Ciani et al.\ (1984) modeled the IUE spectrum
plus $U$, $B$, and $V$ photometry of the M33 nucleus.  They found that
the data were best represented by two stellar populations, one
extremely young ($\sim 10^{7}$ years) and one old ($\sim 10^{10}$
years).  They argued that the effects of foreground and internal dust
were well represented by a screen of MW-type dust
with an $E(B-V) = 0.06$.  Schmidt et al.\ (1990) obtained a long-slit
optical spectrum of the M33 nucleus with a $3\farcs 5$ slit and
modeled it using a combination of star clusters of different ages.
They found that the M33 nucleus was composed of 7 star clusters with
ages ranging from $10^{7}$ to $> 10^{10}$ years.  This is similar to other
studies where over 50\% of the V band light was claimed to be from stars
with ages $> 5 \times 10^{9}$ years.  In addition, Schmidt et al.\
(1990) modeled the effects of dust as due to a Galactic foreground
screen with $E(B-V)_G = 0.05$ and a M33 internal screen with $E(B-V)_i
= 0.05$.  The mass-to-light ratio ($M/L_V$) we compute from our model
fits, corrected for foreground and internal dust, is $\sim$0.1.  This
is significantly different from previous studies, which found $M/L_V
\sim$ 2--3 (\cite{gal82}; \cite{oco83}) which is in excess of the
value likely to be present from dynamical considerations. By including
dust properly, we have arrived at a simpler model of the M33 nucleus
which reproduces its observed SED from the UV to the near-IR.

  We can estimate the peak luminosity of the M33 starburst, although
there is some uncertainty because our models do not constrain the rate
of star formation with time uniquely. For example, the starburst mass
in M33 is roughly 500 times less than that in M82, assuming similar
IMFs (e.g., IMF8 of \cite{rie93}) and the somewhat extended burst
which best fits the M82 data. Thus, at an age of $\sim$10 million
years, the M33 starburst would have luminosity of 0.002 times that of
M82, or $\sim 10^8 L_{\sun}$, Alternately, from our abrupt burst model
and the current luminosity of $1.4 \times 10^7 L_{\sun}$, the
luminosity at 10 Myrs would have been $2.4 \times 10^8 L_{\sun}$.  The
center of the Milky Way contains a very compact ($r \sim 0.2$ pc)
cluster of young ($\le 10$ Myrs old) stars with an integrated
luminosity of $\sim$$10^7 L_{\sun}$ (\cite{rie82}; \cite{wer89}).  The
analogy between these two cases is striking. It appears that 
in M33 we may well be seeing a later development of an event virtually
identical to the starburst that currently dominates the Galactic
Center.

  The relatively young age of the nucleus that we derive from our
models may clear up another mystery.  The M33 nucleus has a very
strong X-ray source with a $L_X \sim 10^{39}$ ergs s$^{-1}$
(\cite{tri88}; \cite{sch95}).  The detection of a 106 day period
requires that this object be a single source (\cite{dub97}).  The only
known type of object which could reproduce both observed effects is a
stellar mass black hole ($\sim$$10 M_{\sun}$) accreting from a
companion star.  This type of 
object was mentioned by Dubus et al.\ (1997), but discarded as the
solution as previous work on the M33 nucleus argued against a
significant population of O--B stars.  The requirements of such an
X-ray binary are a black hole with a companion star with enough mass
to withstand the supernova blast that created the black hole without
being ejected from the binary.  For example, a black hole with an
early B giant companion would be transfering a significant amount of
matter to the black hole, resulting in the observed X-ray luminosity.

\subsection{Nature of the Dust}

\begin{figure}[tbp]
\begin{center}
\plotone{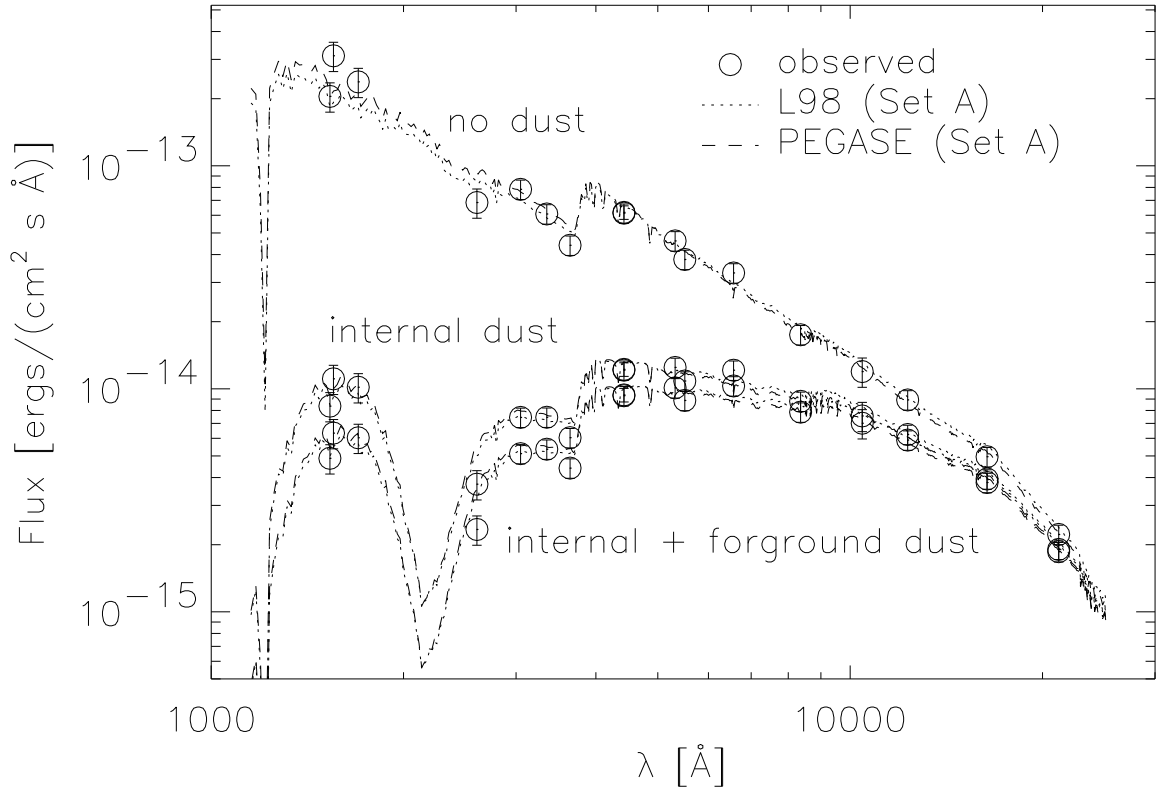}
\caption{The photometric SED and two best fit model SEDs in Set A are
plotted as observed (bottom), corrected for foreground Galactic dust
($\tau_V = 0.20$, middle), and corrected for both foreground and
internal dust (top).  The bottom set of open circles are the observed
data.  The middle and top set of open circles are what the observed
data would look like corrected for foreground Galactic dust and for
both internal and foreground dust.
\label{fig_dust_ex}}
\end{center}
\end{figure}

  The type, geometry, and amount of dust associated with the M33
nucleus is fairly well constrained by the observed photometric SED.
The dust is best described by a shell of somewhat clumpy MW-like dust
($k_2/k_1 \sim 0.5$) with a radial V-band optical depth of 
$\sim$2.  To show the effects of both foreground and internal
dust on the intrinsic SED of the M33 nucleus, we present
Figure~\ref{fig_dust_ex}.  Even in the Galactic foreground dust
corrected SED, the flux at 2607~\AA\ (UIT-A1) is significantly lower
than the flux in the bands around 1600~\AA\ (UIT-B1, F160BW, and
F170W) and around 3300~\AA\ (F300W, F336W, U).  The only known way to
get such a depression in the SED of a stellar population is to
attenuate its SED by dust with a 2175~\AA\ feature, i.e.\ MW-type dust.

  It is interesting to compare this with the result found by Gordon e
al.\ (1997).  They found that SMC-like dust was responsible for
the attenuation in the 30 starburst galaxies in their sample.  They
used IUE-selected starburst galaxies, which biased their study towards
intrinsically bright, lightly attenuated starburst regions.  The
interstellar environment near an intense starburst is very harsh,
associated with an intense UV radiation density and supernovae shocks.
There is evidence from the observed dust extinction curves in the
Magellanic Clouds which suggests that star formation modifies nearby
dust.  In the SMC, the only extinction curve outside the star forming
bar is very similar to that found in the Milky Way.  The other three
extinction curves in the bar are the most extreme known in terms of a
non-existent 2175~\AA\ bump and strong far-UV rise (\cite{gor98}).  In
the LMC, there are two different extinction curves (\cite{mis98}).
The extinction curve associated with stars near the LMC 2 supergiant
shell (which is itself near the 30 Dor star formation region) has a
weak 2175~\AA\ bump.  The extinction curve associated with stars
elsewhere in the LMC is similar to that found in the Milky
Way. However, the 30 Dor region in the LMC exhibits much more intense
star formation than any region in the SMC, but the most extreme
extinction curves are found in the SMC.  This fact implies that the
relationship between star formation activity and the modification of
nearby dust is not a simple one.

  Recent theoretical models of dust exposed to shocks (\cite{jon96};
\cite{odo97}) illustrate the complicated behavior expected for dust
near active star formation.  The number of small particles in shocked
dust is increased through shattering of larger grains and results in
an increased extinction in the far-UV.  However, the models which
produce a stronger far-UV rise, also produce normal or strong
2175~\AA\ bumps.  This is not what is seen in the behavior of dust in
starburst galaxies or the Magellanic clouds and probably reflects our
incomplete understanding of the carrier of the 2175~\AA\ feature.

  The M33 nucleus was also observed by IUE, but is not typical of the
IUE sample used by Gordon et al.\ (1997).  M33 is much closer than the
majority of the IUE sample (0.795 versus 60 Mpc) and its faintness
implies it is intrinsically fainter and/or suffers a higher
attenuation than the rest of the sample.  In the case of M33, the
starburst is surrounded by a larger optical depth of dust and, since
it lacks emission lines, is older than those seen in the IUE sample.
Thus, the dust in the M33 nucleus has likely undergone less processing
since it is 
near a small starburst and self-shielded by a large amount of dust.
The amount by which dust can be processed in starburst regions is
likely related to the mass and age of the starburst as well as the
extent to which the dust can shield itself.  Additional work on other
fairly reddened starburst regions is needed to test this
interpretation of the origin of the MW-type dust we find in the M33
nucleus.

\subsection{Possible Role of Collisions and Mergers}

  The low velocity dispersion and extreme compactness of the M33
nucleus have caused Lauer et al.\ (1998) to suggest that stellar
mergers might play an important role in creating the blue stellar
population 
(see also \cite{kor93}).  These arguments are based on a calculation
that a typical star lying within the present-day environment of the
M33 nucleus would have a collision/binary capture timescale of about
a Hubble time.  This model could apply if the previous estimate of an
old ($10^{10}$ yr) stellar population in the nucleus of M33 is assumed
(\cite{cia84}; \cite{sch90}).  That is, if conditions in the nucleus have
not changed over a Hubble time, most stars in the nucleus will have
undergone such an event.  To assess if collisions and mergers
are important, we will assume the age of the nucleus is on the order
of the Hubble time and see if reasonable arguments lead to the
observed characteristics of the M33 nucleus.

  The optical spectrum of M33 (Figure~\ref{fig_id_opt}) demonstrates
that the nucleus contains many stars of $\sim 2.5 - 6 ~M_{\sun}$.  The
observed nuclear B magnitude is 14.6.  Corrected for foreground and
internal dust, the intrinsic B magnitude is 12.5, corresponding to a
M$_{\rm B} = -12.0$ (for a distance of 795 kpc).  For the purposes of
this exercise, we attribute most of
the nuclear mass of $5 \times 10^5~M_{\odot}$ to an old stellar
population that creates more massive stars by mergers.  The absolute
blue magnitude would require the presence of $\sim 10^5$ main sequence
stars with masses $\sim$3 $M_{\sun}$, if these stars were the most massive 
in the nucleus.  This hypothetical population accounts for an
uncomfortably large portion of the total nuclear mass.  Instead, it
seems required that the light from the nucleus is dominated by stars
closer to the top of the permitted mass range ($6 M_{\sun}$).
We assume the old stellar population is described by an IMF similar to 
the one taken for the 
starburst, up to a main sequence turnoff mass of $0.9~M_{\sun}$.
Allowing for the mass lost in mergers, stars at the turnoff must
undergo on the order of six to eight mergers to build a $5 -
6~M_{\sun}$ star.  Thus, the calculation by Lauer et al.\ (1998) that,
on average, each star might undergo one merger in a Hubble time
implies that the dominant stars could not be created in this
fashion.

  A more rigorous argument would account for the dynamical evolution
of the nucleus.  If the age of the M33 nucleus is on the order of the
Hubble time, then it is possible that it has gone through core
collapse (\cite{her91}; \cite{kor93}; \cite{lau98}).  Lee (1987) has
studied stellar mergers in dynamical models of systems of $0.7
~M_{\sun}$ stars which, in their post core collapse configuration,
resemble the M33 nucleus.  The merger rate reaches a sharp peak around
core collapse.  Stars that are the result of eight mergers, as
required to build the blue stars in the M33 nucleus, are $10^{-5}$
less common than the original population members shortly after core
collapse in Lee's (1987) model.  We have roughly corrected this result
for the approximately three times higher number of seed stars in M33 compared
with the modeled cluster.  In addition, Lee (1987) made the optimistic
assumptions that all binary captures lead to mergers (rather than the
expected $\sim$25\%; \cite{lee86}) and that the merged stars have
main sequence lifetimes similar to normal stars (whereas the lifetimes
are shortened considerably by the presence of enriched nuclear
material; e.g., \cite{sil97}). We project that mergers would yield at
best an order of magnitude less than the numbers of massive stars
required. This conclusion agrees with the statement by Hernquist et
al.\ (1991).

  The merger hypothesis is attractive primarily because of the very
small observed core radius in M33.  However, the observations are
dominated by the relatively massive stars; hence we have to examine
the time scale for mass segregation to be sure that the radius
measured is representative of the distribution of lower mass stars
that hypothetically merge due to their high density in this core.  We
have estimated the dynamical decay time for stars of $5 ~M_{\sun}$ in a
total mass of $1 \times 10^6 ~M_{\sun}$ distributed over a radius of
0.5 pc, assuming that the density goes as $1/r^2$, $1/r$, or is
constant within this radius.  The calculations are based on equation
(7-18) of Binney \& Tremaine (1987).  We find that the dynamical decay
time from formation at a radius of 0.5 pc is in all three cases
substantially less than the lifetimes of the stars; that is, mass
segregation is likely to have occurred.  The increasing size of the
nucleus with wavelength (see Table~\ref{tab_nuc_size}) and
corresponding observed color gradients (\cite{kor93}; \cite{lau98})
confirm that the blue, massive stars have probably sunk to the center
and are surrounded by a population of lower mass stars.  Thus,
estimates of the merger rate based on the small observed core radius
(e.g., \cite{lau98} and our discussion in the preceding paragraphs)
are likely to be too high.

  The merger hypothesis can also be tested by examining the properties
of the merger products.  These products have huge angular momentum
compared with normal stars (e.g., \cite{ras95}). If the
interiors become convective, the resulting high rotation rates may be
spun down through magnetic interaction with circumstellar disks
(\cite{leo95}).  However, the resulting mixing of metal rich
material causes the evolution of the stars to differ significantly
from that of normal stars (e.g., \cite{bai95}; \cite{sil97}). The
possibility of abnormal stellar evolution is made very 
unlikely by the success in our spectral synthesis evolutionary models
in fitting the properties of the nucleus across the entire spectrum,
since these models assume normal evolution.  The possibility that the
nuclear population is characterized by abnormally high rotational
velocities is contradicted by the small net velocity dispersion.

  If the merger mechanism played an important role in M33, then the
conclusions drawn from starburst models would largely be
vitiated.  However, a variety of arguments show that mergers play a
negligible role and hence that the blue stellar population must arise
in a starburst as described in Section \ref{sec_stell_pop}.

\section{Conclusions}

  By fitting the UV to near-IR SED of the M33 nucleus with a starburst 
model which includes stars, gas, and dust, we find the M33 nucleus to
be composed 
of a $\sim$70 Myr burst of star formation with a mass of $\sim$$7
\times 10^{5} ~M_{\sun}$ enshrouded by a shell of MW-type dust
($\tau_V \sim 2$).  This result is different from previous work which
modeled the nucleus using both young ($10^7 - 10^8$ years) and old
($10^9 - 10^{10}$ years) stellar populations subject only to modest
reddening.  By including the effects of dust internal to M33, we are
able to reproduce the relatively red SED of the M33 nucleus without
requiring a luminous old stellar population.  As a result, this work
is more consistent with the small size and the detection of far-UV
flux from the nucleus.

  The dust in M33 has a strong 2175~\AA\ bump, in contrast to other
UV-bright starbursts that do not show this feature. Not much can be
said with a sample of one galaxy, but it is possible that the
different dust characteristics in M33 nucleus are related to the
higher optical depth of dust and/or the age of the starburst. Thus,
there is not a single type of dust found in starburst galaxies but a
range of types varying from that found in the MW to the SMC.
Correcting for 
the effects of dust in starburst galaxies must be done with care,
especially in the UV.

 The M33 nuclear starburst resembles the ultracompact population of
very massive and luminous stars in the center of the Milky Way, both
in the peak luminosity deduced for the starbursts responsible and in
the size of the starburst region. It seems likely that ultracompact
starbursts ($< 1$ pc) are common in the nuclei of normal galaxies and
may influence the evolution of these regions significantly.

  Confirmation of our interpretation of the M33 nucleus would be greatly
helped by additional UV and far-IR observations.  A good UV spectrum
of just the nucleus would be easily obtained by the STIS instrument on
HST due to it's high sensitivity and small apertures.  Far-IR
observations at high spatial resolution would help put stringent
constraints on the mass of the dust internal to the nucleus.  These
observations should be possible with SIRTF.

\acknowledgements

  We thank the anonymous referee for valuable suggestions which
improved the presentation of our work.  This work benefited from
discussions about SES models with Claus Leitherer, Michel Fioc, and
Brigette Rocca-Volmerange.  In addition, we are indebted to Claus
Leitherer and Michel Fioc for providing us with new grids of SEDS from
the L98 and PEGASE SES models.  We thank Juhan Frank for helpful
discussion on the X-ray properties of binaries and the intricacies of
cluster dynamics.  Support for this work was provided by NASA through
grant number [AR-08002.01-96A] from the Space Telescope Science
Institute. Additional support was provided by the NSF under grant
AST95-29190.  MMH received support for this work provided by NASA
through Hubble Fellowship grant \#HF-1072.01-94A awarded by the Space
Telescope Science Institute.

\end{document}